\documentclass[aps,prl,showpacs,twocolumn]{revtex4}
\usepackage{amsfonts}
\usepackage{amsmath}
\usepackage{amssymb}
\usepackage{epsfig,graphicx,amssymb}

\input{.tex}

\begin{document}

\title{Topological Force and Torque in Spin-Orbit Coupling System }
\author{F. J. Huang$^{1,2}$, R. Qi$^{1}$, Y. D. Li$^{2}$, W. M. Liu$^{1}$}
\address{$^1$Beijing National Laboratory for Condensed Matter Physics,
Institute of Physics, Chinese Academy of Sciences, Beijing 100080,
China}
\address{$^2$School of Physics, Yunnan University, Kunming 650091,
China}

\date{\today}

\begin{abstract}
The topological force and torque are investigated in the systems
with spin-orbit coupling. It is demonstrated that the topological
force and torque appears as a pure relativistic quantum effect in
an electromagnetic field. The origin of both topological force and
torque is the $Zitterbewegung$ effect. Considering nonlinear
behaviors of spin-orbit coupling, we address possible novel
phenomena driven by the topological forces.
\end{abstract}


\maketitle

Recently, the spin-orbit coupling has become an interesting topic
due to the spin Hall effect [1-5]. It provides an efficient route
to generate and control quantum spin state electrically. Generally
speaking, the spin-orbit coupling arises as relativistic quantum
effect from the Dirac equation, and describes the interaction of
the electron spin, momentum and electromagnetic field. In the
system with spin-orbit coupling, the semiclassical equations of
electron were studied recently, and some novel effects have been
found [6-15]. However, in the systems with spin-orbit coupling,
there remain some questions to be answered. As well known, the
non-relativistic approximation of Dirac equation can be obtained
from Foldy-Wouthuysen transformation. In the transformation, there
exists a gauge potential in momentum space [15,16], so one
question is, what is the position operator in semiclassical
equations? Since the position operator is the space-time
parameter, its definition is significant. If we are uncapable of
defining it correctly, we couldn't have the complete understanding
for spin-orbit coupling.

In this paper, we investigate the gauge field of position operator
in momentum space, and its related effects. Based on the dynamic
continuity equation, we derive the quantum force and torque which
contain two parts. The conventional part has the same form as in the
classic electrodynamics, and the topological part originates from
the spin-orbit coupling and other terms from relativistic quantum
correction. We notice that the topological part has a close relation
to the $Zitterbewegung$ effect. For a two-dimensional system in a
magnetic field, we propose that the topological force and torque can
reveal more complicated phenomena.

\textbf{Topological velocity} The Dirac equation of electron with the wave
function $\Psi =(\varphi, \chi)^{T}$ reads $i\hbar \frac{\partial }{%
\partial {t}}\Psi =H\Psi $, with the Hamiltonian $H=c\boldsymbol{\alpha}%
\cdot \boldsymbol{\pi}+\beta mc^{2}+V(\mathbf{x})$, where $\boldsymbol{\pi}=%
\mathbf{P}-\frac{e}{c}\mathbf{A}$, $\boldsymbol{\alpha}$ and $\beta $ are
the $4\times 4$ Dirac matrices, $V(\mathbf{x})=e\phi $ is a scalar
potential, and $\mathbf{A}$ is a vector potential for a magnetic field, $%
\mathbf{B}=\nabla \times \mathbf{A}$. $m$ and $e$ are the electron mass and
charge, and $c$ is the speed of light. To reveal the spin-orbit coupling in
Dirac equation, we perform the Foldy-Wouthuysen transformation $\Psi
^{^{\prime }}=U(\boldsymbol{\pi})\Psi $, where $U(\boldsymbol{\pi})=e^{iS}$,
it is a unitary transformation [17,18]. Choosing $S=-i(\beta %
\boldsymbol{\alpha}\cdot \boldsymbol{\pi}/2mc)$, and substituting it into
the Dirac equation, we obtain the transformed Hamiltonian
\begin{equation}
H^{^{\prime}}= \beta mc^{2}+e^{iS}c\boldsymbol{\alpha}\cdot \boldsymbol{\pi}%
e^{-iS}+e^{iS}V(i\hbar \partial _{\mathbf{p}})e^{-iS}-e^{iS}i\hbar \partial
_{t}e^{-iS}.
\end{equation}
The scalar potential becomes $V(\mathbf{D})$ with covariant derivative
defined by $\mathbf{D}=i\hbar \partial _{\mathbf{p}}+\mathcal{A}$, with the
pure gauge potential $\mathcal{A}=i\hbar U(\boldsymbol{\pi})^{\dagger
}\partial _{\mathbf{p}}U(\boldsymbol{\pi})$. By defining the covariant
derivative, we have the position operator $\mathbf{X}=i\hbar \partial _{%
\mathbf{p}}+\mathcal{A}$. The gauge potential $\mathcal{A}$ here is trivial.
Neglecting the inter-band transition, and considering the adiabatic
approximation, we find the non-trivial gauge potential
\begin{equation}
\mathcal{A}=\frac{\hbar (\boldsymbol{\pi}\times \Sigma )}{4m^{2}c^{2}},
\end{equation}%
where $\Sigma =1\otimes \boldsymbol{\sigma}$.

By expanding the Hamiltonian $H^{^{\prime }}$, and taking $S^{^{\prime
}}=\hbar e\boldsymbol{\alpha}\cdot \mathbf{E}/4m^{2}c^{3}$ to do another
transformation $\Psi ^{^{\prime \prime }}=e^{iS^{^{\prime }}}\Psi ^{^{\prime
}}$, we get the non-relativist Hamiltonian
\begin{equation}
\begin{split}
H_{sch}=& \frac{\boldsymbol{\pi}^{2}}{2m}+e\phi -\frac{e\hbar }{2mc}%
\boldsymbol{\sigma}\cdot \mathbf{B}-\frac{\hbar ^{2}e}{8m^{2}c^{2}}\nabla
\cdot \mathbf{E} \\
& -\frac{e\hbar }{8m^{2}c^{2}}\boldsymbol{\sigma}\cdot (\mathbf{E}\times %
\boldsymbol{\pi}-\boldsymbol{\pi}\times \mathbf{E}),
\end{split}%
\end{equation}%
where the first two terms refer to kinetic energy and scalar
potential, the third is the Zeeman energy, the fourth term is a
Darwin term, refers to the correction of scalar potential, and the
fifth term is so-called the spin-orbit coupling term. With this
Hamiltonian, the Schr\"{o}dinger equation
with wave function $\psi $ then can be written as $i\hbar \frac{\partial }{%
\partial {t}}\psi =H_{sch}\psi $, where $\psi =(1+(\boldsymbol{\sigma}\cdot %
\boldsymbol{\pi})^{2}/8m^{2}c^{2})\varphi $ is the wave function
$\varphi $ which have been performed unitary transformations. After
the unitary transformations, the wave function $\chi \sim 1/c^{3}$,
so that we can neglect it. Based on this fact, we can regard the
wave function $\psi $ have been done the normalization. In the
Schr\"{o}dinger equation of $\psi $, the position operator has been
redefined by a covariant derivative in the
momentum space $\mathbf{X}=\mathbf{D}=i\hbar \partial _{\mathbf{p}}+\mathcal{%
A}$, with the gauge potential $\mathcal{A}=\hbar (\boldsymbol{\pi}\times %
\boldsymbol{\sigma})/4m^{2}c^{2}$.

We now turn our attention to the gauge potential induced by transformation $%
U(\boldsymbol{\pi})$. In our process, we consider the adiabatic
approximation, so it is straightforward to interpret the gauge potential $%
\mathcal{A}$ as the non-Abelian Berry gauge potential, and it is Berry
connection in momentum space, the Berry phase $\theta =\oint_{C}d\mathbf{%
p\cdot }\mathcal{A}$, where $C$ is loop in the momentum space. The Berry
gauge potential describes the non-trivial geometry of the fiber bundle of
the Hamiltonian eigenvectors over the phase space, it describes the
influence of negative energy states in positive energy space.

Using $\mathcal{A}=\hbar (\boldsymbol{\pi}\times \boldsymbol{\sigma}%
)/4m^{2}c^{2}$, and the relation $F_{ij}=\partial _{p_{i}}\mathcal{A}%
_{j}-\partial _{p_{j}}\mathcal{A}_{i}-(i/\hbar)[\mathcal{A}_{i},%
\mathcal{A}_{j}]$, we get the non-Abelian Berry gauge curvature $%
F_{ij}=-\varepsilon _{ijk}\hbar (\sigma _{k}/2m^{2}c^{2}-(\boldsymbol{\sigma}%
\cdot \boldsymbol{\pi})\pi _{k}/8m^{4}c^{4})$. Let $\lambda =-\frac{\hbar }{2%
}\sigma _{k}$, $\boldsymbol{\pi}\approx \mathbf{P}$, introducing a
dual vector $F_{k}=\varepsilon _{ijk}F_{ij}$, then the gauge field
can be rewritten as $F_{k}=(\lambda
/m^{2}c^{2})(1-\mathbf{P}^{2}/4m^{2}c^{2})$, where the first term
presents a monopole taking the radial length at the rest momentum
$mc$ with its strength given by $\frac{\hbar }{2}\sigma _{k}$ in the
momentum space, the second term is the higher order of correction.
If we neglect the second term in $F_{k}$, then
\begin{equation}
F_{k}=\frac{\lambda }{(mc)^{2}}.
\end{equation}%
We note that this field only has two values $\pm \frac{\hbar }{2}$ in the
momentum space, so the difference between this field and the ordinary field $%
F_{k}=\lambda /\mathbf{P}^{2}$ [19], is that this field is uniform.
Obviously, this gauge field is a consequence of non-relativistic
approximation $\mathbf{P}\ll mc$. In classical dynamics, the rest energy $%
mc^{2}$ does not affect the motion of particle in non-relativistic limit.
However, in quantum mechanics, the rest momentum $mc$ induces a gauge field
in the momentum space, then it will influence the motion of particle.

We notice that this gauge field can induce topological effects.
Noting that after the unitary transformation, the covariant position
operator of Dirac equation is defined by $\mathbf{X}$, and the
commutator of covariant operator is not trivial:
$[x_{i},x_{j}]=i\hbar F_{ij}$. This relation is characteristic of
the unitary transformation, it is this relation that
induces the gauge field in non-relativistic approximation. Defining $\mathbf{%
x}=i\hbar \partial _{\mathbf{p}}$ as the canonical position operator, then $%
\mathbf{X}=\mathbf{x}+\mathcal{A}$, and the semiclassical equation of motion
is $\dot{\mathbf{X}}=\dot{\mathbf{x}}+\dot{\boldsymbol{\pi}}\times \mathcal{F%
}$, with $\mathcal{F}_{k}=F_{k}/2$. Apparently, the last term of the
equation above is a topological term, and results from the non-trivial
commutator of covariant position operator. It acts as the Lorentz force in
the momentum space.

By using the Hamiltonian (3), we define the velocity $\dot{\mathbf{X}}=\dot{%
\mathbf{x}}+(e\hbar /4m^{2}c^{2})(\mathbf{E}+\frac{\boldsymbol{\pi}}{mc}%
\times \mathbf{B})\times \boldsymbol{\sigma}$, where $\dot{\mathbf{x}}=%
\boldsymbol{\pi}/m+(e\hbar /4m^{2}c^{2})\mathbf{E}\times \boldsymbol{\sigma}$
is the conventional velocity. The second term in $\dot{\mathbf{X}}$ actually
is a topological term, it is caused by the gauge field $F_{k}=\lambda
/(mc)^{2}$, and we regard it as a topological velocity. Introducing the
electron intrinsic magnetic moment $\hat{\mathbf{m}}=(e\hbar /2mc)%
\boldsymbol{\sigma}$, we can rewrite the velocity as
\begin{equation}
\mathbf{v}=\tilde{\mathbf{v}}+\frac{(\mathbf{E}^{^{\prime }}\times \hat{%
\mathbf{m}})/2c}{m},
\end{equation}%
where $\tilde{\mathbf{v}}=\dot{\mathbf{x}}$ is the conventional velocity
operator which corresponds to position operator $\mathbf{x}$, $\mathbf{E}%
^{^{\prime }}=\mathbf{E}+(\boldsymbol{\pi}/mc)\times \mathbf{B}$ is
the total electric field which affects an electron in its local
coordinate frame. If we remind that the gauge field is induced by
the interference between the negative and positive energy states,
this topological term in Eq. (5) is contributed by the rest energy
under the influence of negative states. Comparing with two
velocities, we find the terms concern with $\mathbf{E}$ in Eq. (5)
is twice that of conventional velocity. Besides, our definition adds
a contribution of magnetic field. One can easily find that these
differences between the two velocities is caused by the gauge field
in momentum space.

\textbf{Topological Force and Torque} It is very interesting in the
topological force which acts on the electron in an electromagnetic field.
Let us start from the dynamic continuity equation. The dynamic continuity
equation represents the change form of a physical quantify in space-time, it
always requires the existence of a source such as a force or a torque.

Denote by $\mathbf{\rho }^{M}=\psi ^{\dagger }\hat{\mathbf{\rho }}^{M}\psi $
the momentum density, where $\hat{\mathbf{\rho }}^{M}=m\mathbf{v}$. Using $%
\frac{\partial }{\partial t}\psi ^{\dagger }\hat{\mathbf{\rho }}^{M}\psi =%
\dot{\psi}^{\dagger }\hat{\mathbf{\rho }}^{M}\psi +\psi ^{\dagger }\hat{%
\mathbf{\rho }}^{M}\dot{\psi}$ and $i\hbar \dot{\psi}=H_{sch}\psi $, then we
arrive at
\begin{equation}
\mathbf{\dot{\rho}}^{M}+\nabla \cdot \mathbf{J}^{M}=\mathbf{f},
\end{equation}%
where $\mathbf{J}^{M}=\psi ^{\dagger }\hat{\mathbf{J}}^{M}\psi =(1/2)\psi
^{\dagger }\{\hat{\mathbf{\rho }}^{M},\tilde{\mathbf{v}}\}\psi $, and $%
\tilde{\mathbf{v}}$ is the conventional velocity. The force is given by $%
\mathbf{f}=(1/i\hbar )[\mathbf{\rho }^{M},H_{sch}]=\mathbf{f}_{con}+\mathbf{f%
}_{top}$, where $\mathbf{f}_{con}$ is the conventional force which
corresponds to the position operator $\mathbf{x}$, and $\mathbf{f}_{top}$ is
the topological force. The conventional force can be read $\mathbf{f}_{con}=%
\mathbf{\rho }^{C}\mathbf{E}+(\mathbf{J}^{C}\times \mathbf{B})/c+(\nabla
\mathbf{B})\cdot \mathbf{m}+(\nabla \mathbf{E})\cdot \mathbf{P}+(e/2mc^{2})%
\mathbf{E}\times (\hat{\mathbf{m}}\times \mathbf{B}^{^{\prime }})$, where $%
\rho ^{C}=e\psi ^{\dagger }\psi $ and $\mathbf{J}^{C}=e\psi ^{\dagger }%
\tilde{\mathbf{v}}\psi $ are the charge and the current densities for $%
\mathbf{f}_{con}$. $\mathbf{m}=\psi ^{\dagger }\hat{\mathbf{m}}\psi $, and $%
\mathbf{P}$ is defined by $\mathbf{P}=\psi ^{\dagger }\hat{\mathbf{P}}\psi
=\psi ^{\dagger }(\frac{\boldsymbol{\pi}}{2mc}\times \hat{\mathbf{m}})\psi $%
, it correlates with the electrical polarization induced by the spin-orbit
coupling [10]. $\mathbf{B}^{^{\prime }}=\mathbf{B}-(\boldsymbol{\pi}%
/2mc)\times \mathbf{E}$ presents the total magnetic field which an
electron experiences in its local coordinate frame. Meanwhile,
considering the topological velocity, the topological force can be
found to be
\begin{equation}
\mathbf{f}_{top}=\frac{e}{2mc^{2}}\mathbf{E}^{^{\prime }}\times (\hat{%
\mathbf{m}}\times \mathbf{B}^{^{\prime }}).
\end{equation}

It is clear that the topological force is an additional term to the
conventional force, so that we can divide the total force into two parts: $%
\mathbf{f}=\mathbf{f}_{cla}+\mathbf{f}_{so}$ where $\mathbf{f}_{cla}$ is the
force the electron experiencing in classical motion, and $\mathbf{f}_{so}$
is the force produced by the spin-orbit coupling. The force $\mathbf{f}_{cla}
$ can be written as $\mathbf{f}_{cla}=\mathbf{\rho }^{C}\mathbf{E}+(\mathbf{J}%
^{C}\times \mathbf{B})/c+(\nabla \mathbf{B})\cdot \mathbf{m}+(\nabla \mathbf{E}%
)\cdot \mathbf{P}$, it is well known that the classical forces are Lorentz,
\textquotedblleft Stern-Gerlach", and electric dipole forces. All the forces
above have correspondence forces in classical dynamics, so we combine them
to a classical force $\mathbf{f}_{cla}$.

The force $\mathbf{f}_{so}$ can be read as $\mathbf{f}_{so}=(e/mc^{2})%
\mathbf{E}^{^{\prime \prime }}\times (\hat{\mathbf{m}}\times \mathbf{B}%
^{^{\prime }})$, where $\mathbf{E}^{^{\prime \prime }}=\mathbf{E}+(%
\boldsymbol{\pi}/2mc)\times \mathbf{B}$. To explore the physical meaning, we
divide this force into two parts, $\mathbf{f}_{so}=e\psi ^{\dagger }\frac{(%
\mathbf{E}^{^{\prime \prime }}\times \hat{\mathbf{m}})}{mc^{2}}%
\times \mathbf{B}^{^{\prime }}\psi +e\psi ^{\dagger }\frac{(\mathbf{B%
}^{^{\prime }}\times \mathbf{E}^{^{\prime \prime }})}{mc^{2}}%
\times \hat{\mathbf{m}}\psi $, where the first term indicates the
electron experiencing an electrical field $\mathbf{E}^{^{\prime
\prime
\prime }}=[(\mathbf{E}{^{\prime \prime }}\times \hat{\mathbf{m}})/mc^{2}]\times \mathbf{B}%
^{^{\prime }}$, and it is an electric force. One can find that the
second term of $\mathbf{f}_{so}$ is a pure quantum effect, since
$(\mathbf{E}\times \mathbf{B}^{^{\prime }})/mc^{2}$ is
dimensionless, this term does not concern with any classical
quantify, it is a pure quantum force, as shown in Fig. 1.

When the magnetic field vanishes, the force $\mathbf{f}_{so}$ reduces to $%
\mathbf{f}_{ele}=-\psi ^{\dagger }(e/2mc^{2})(\hat{\mathbf{m}}\cdot \mathbf{E%
})(\frac{\boldsymbol{\pi}}{mc}\times \mathbf{E})\psi $. This force
originates from the second term of $\mathbf{f}_{so}$, so it is a pure
quantum force, it can be used to explain the $Zitterbewegung$ of the
electron in spin-orbit coupling system [7,8]. The direction of this force is
perpendicular to the electrical field, and to the velocity and the spin
polarization direction, so it is a transverse force. When the electrical
field vanishes, the force $\mathbf{f}_{so}$ becomes $\mathbf{f}_{mag}=-\psi
^{\dagger }(e/2mc^{2})\hat{\mathbf{m}}\cdot (\frac{\boldsymbol{\pi}}{mc}%
\times \mathbf{B})\mathbf{B}\psi $, this force also originates from the
second term of $\mathbf{f}_{so}$, and it also is a pure quantum force. The
direction of this force parallels with the magnetic field, and is
perpendicular to the velocity and the spin polarization direction, so this
force also is a transverse force. In fact, this force has duality with $%
\mathbf{f}_{ele}$, so it also can be known from the $Zitterbewegung$ effect.

\begin{figure}[tbp]
\begin{picture}(0,60)(120,0)
\thicklines \put(0,0){\line(3,2){72}} \put(160,0){\line(3,2){72}} 
\put(0,0){\line(1,0){160}} \put(72,48){\line(1,0){160}}
\put(0,-2){\line(0,1){2}} \put(160,-2){\line(0,1){2}}
\put(232,45){\line(0,1){3}}
\put(65,27){\vector(0,1){35}}
\put(85,40){\vector(-3,-2){50}}
\put(0,-2){\line(1,0){160}}
\put(160,-2){\line(3,2){72}} 
\put(80,12){\vector(3,2){15}}
\put(175,40){\vector(-3,-2){15}}
\put(70,12){\vector(1,0){80}}\put(110,40){\vector(1,0){80}}
\put(97,20){$\boldsymbol{f}_{so}$} \put(185,30){$J^{C}_{\uparrow}$}
\put(52,53){$\emph{\textbf{B}}$} \put(48,7){$\emph{\textbf{E}}$}
\put(134,18){$J^{C}_{\downarrow}$}
\put(150,26){$\boldsymbol{f}_{so}$}
\end{picture}
\caption{The electrical field $\mathbf{E}$ and the magnetic field
$\mathbf{B}$ exert the pure quantum forces $\boldsymbol{f}_{so}$
on the opposite spin polarized electrons, where $J^{C}_{\uparrow}$
and $J^{C}_{\downarrow}$ are charge currents with different spin
polarization.}
\end{figure}
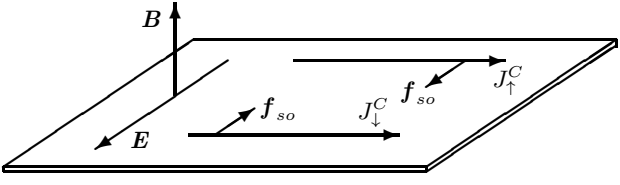

We now turn to the angular momentum continuity equation. Denoting $\mathbf{%
\rho }^{AM}$ as the total angular momentum density, then $\mathbf{\rho }%
^{AM}=\mathbf{\rho }^{\mathcal{O}}+\mathbf{\rho }^{S}$, where the orbit
momentum density $\mathbf{\rho }^{\mathcal{O}}=\psi ^{\dagger }\mathbf{X}%
\times \hat{\mathbf{\rho }}^{M}\psi $, the spin density $\mathbf{\rho }%
^{S}=(\hbar /2)\psi ^{\dagger }\boldsymbol{\sigma}\psi $. Similar to the
momentum continuity equation, we arrive at
\begin{equation}
\mathbf{\dot{\rho}}^{AM}+\nabla \cdot \mathbf{J}^{AM}=\mathbf{T}.
\end{equation}%
The angular momentum current is defined by $\mathbf{J}^{AM}=(1/2)\psi
^{\dagger }\{\mathbf{X}\times \hat{\mathbf{J}}^{M}+(\hbar /2)%
\boldsymbol{\sigma},\tilde{\mathbf{v}}\}\psi $, where the first term is the
orbit angular momentum current, and the second term is the spin current $%
\mathbf{J}^{^{\prime }S}$. The total torque is given by $\mathbf{T}%
=(1/i\hbar )[\mathbf{\rho }^{AM},H_{sch}]=\mathbf{T}_{con}+\mathbf{T}_{top}$%
, where $\mathbf{T}_{con}$ is the conventional torque which corresponds to
the position operator $\mathbf{x}$, $\mathbf{T}_{top}$ is the topological
torque. The conventional torque can be read $\mathbf{T}_{con}=\psi ^{\dagger
}\mathbf{x}\times \hat{\mathbf{f}}\psi +\mathbf{m}\times \mathbf{B}+\mathbf{P%
}\times \mathbf{E}+\psi ^{\dagger }(\boldsymbol{\pi}/2mc)\times (\mathbf{E}%
\times \hat{\mathbf{m}})\psi $, where $\hat{\mathbf{f}}$ is an
operator with $\mathbf{f}=\psi ^{\dagger }\hat{\mathbf{f}}\psi $.
Meanwhile, considering the topological velocity, the topological
torque reads
\begin{equation}
\mathbf{T}_{top}=\psi ^{\dagger }\mathcal{A}\times \hat{\mathbf{f}}\psi
+\psi ^{\dagger }\mathbf{x}\times \hat{\mathbf{f}}_{top}\psi ,
\end{equation}%
where the first term concerns with the gauge potential, and the second term
contributed by the topological force, so both two terms can be known from
the $Zitterbewegung$ effect.

Of cause, the total torque can be divided into the orbit torque $\mathbf{T}^{%
\mathcal{O}}$ and the spin torque $\mathbf{T}^{S}$. Combining the
conventional and the topological torque, the orbit torque has the form $%
\mathbf{T}^{\mathcal{O}}=\psi ^{\dagger }\mathbf{X}\times \hat{\mathbf{f}}%
\psi $. Obviously, $\mathbf{T}^{\mathcal{O}}$ has a corresponding torque in
classical dynamics. The spin torque is given by $\mathbf{T}^{S}=\mathbf{m}%
\times \mathbf{B}-\psi ^{\dagger }\hat{\mathbf{m}}\times (\frac{%
\boldsymbol{\pi}}{2mc}\times \mathbf{E})\psi $. The first term is
contributed by the magnetic field. The second term is the influence of
electric field $\mathbf{E}$, which can induce a magnetic field in electron's
own coordinate frame, where $1/2$ is Thomas precession factor. These two
terms can be found in Bargman-Michel-Telegdi Equation which describes the
precession of the electron's spin [20].

Now we can write the spin continuity equation as $\dot{\mathbf{\rho}}^{S}
+\nabla\cdot\mathbf{J}^{^{\prime}S}=\mathbf{T}^{S}$, where the spin density
is defined by $\mathbf{\rho}^{S}= (\hbar/2)\psi^{\dagger} \boldsymbol{\sigma}%
\psi$, the spin torque is defined by $\mathbf{T}^{S}$. The spin current is $%
\mathbf{J}^{^{\prime}S}=(\hbar/4)\psi^{\dagger} \{\boldsymbol{\sigma},\tilde{%
\mathbf{v}}\}\psi$, $\tilde{\mathbf{v}}$ is the conventional velocity
operator.

\textbf{Nonlinear Spin-Orbit Coupling} We apply the previous results to a
two-dimensional system under a weak magnetic field $\mathbf{B}=(0,0,B)$,
with $\mathbf{A}=(0,xB,0)$. The Hamiltonian reads $H_{2D}=\pi ^{2}/2m-\mu
_{B}\boldsymbol{\sigma}\cdot \mathbf{B}+(eF/2)(\boldsymbol{\mathbf{E}\times%
\sigma})\cdot \boldsymbol{\pi}+eEx$, where $\mu _{B}$ is Bohr magnetic
momentum, $F=(\hbar /2)/(mc)^{2}$ is the gauge field, the electrical field $%
\mathbf{E}=(E,0,0)$. The topological velocity $v_{y}=\hbar k_{y}/m-\mu
_{B}FB\sigma _{z}k_{y}-eFE\sigma _{z}$, where $k_{y}$ is wave vector. The
spin current operators read $\hat{\mathbf{J}}_{y}^{z}=(\hbar ^{2}/2m)\sigma
_{z}k_{y}-(\hbar /2)\mu _{B}FBk_{y}-(\hbar /2)eFE$, where the superscript
presents the spin direction, the subscript denotes the direction of electron
motion. It is found that the topological term here only concerns with
momentum. In this case, the trajectories read $X=x+\hbar Fk_{y}\sigma _{z}/2$%
, $Y=y$. The spin current
\begin{equation}
\mathbf{J}_{y}^{z}=\sigma _{sH}E,
\end{equation}%
where the spin Hall coefficient $\sigma _{sH}=-(e/4\pi )(\delta
+\delta \epsilon /4-\epsilon /2)$, where $\delta =eBF/c$, $\epsilon
=\hbar k_{F}^{2}F $ are dimensionless parameters, $k_{F}$ is the
Fermi momentum. It is clear that the second term in the spin Hall
coefficient $\sigma _{sH}$ is contributed by the gauge field.

For spin current $\mathbf{J}^{z}_{y}$, there is a pure quantum force
perpendicular to the magnetic field, it is given by $f^{x}_{top}=%
\mu_{B}(eE/mc^{2})\psi^{\dagger}\sigma_{z}\psi B$. We can find that this
force is proportional to the magnetic field. The directions of the spin
polarization, the current and the force are perpendicular to each another.
This force originates from the $Zitterbewegung$ effect.

If we construct a spin-orbit coupling Hamiltonian as follows,
\begin{equation}
H=\frac{\pi ^{2}}{2m}-\mu _{B}\boldsymbol{\sigma}\cdot \mathbf{B}+\eta (%
\boldsymbol{\pi}\times \boldsymbol{\theta})\cdot \boldsymbol{\nu},
\end{equation}%
where $\eta=(\hbar e/4m^{3}c^{3})B$ is a spin-orbit coupling
parameter, $\boldsymbol{\nu}$ is a unit direction which is along the
$z$ direction, and $\boldsymbol{\theta}=
\boldsymbol{\theta}(\boldsymbol{\pi},\boldsymbol{\sigma}
)=\boldsymbol{\pi}\times\boldsymbol{\sigma}$ is a vector presenting
spin-orbit coupling types, we can also obtain the same velocity of
the two-dimensional system above (the system has been eliminated the
scalar potential). This indicates that the Hamiltonian $H_{2D}$,
which takes the gauge field into account, is equivalent to a system
with more complex spin-orbit coupling. This equivalence illuminates
that the gauge field can present more complicated spin-orbit
coupling in spin-orbit coupling system, such as
$\boldsymbol{\pi}\times\boldsymbol{\theta}$, with
$\boldsymbol{\theta}=\boldsymbol{\pi}\times\boldsymbol{\sigma}$ or
$(\boldsymbol{\pi}\cdot\boldsymbol{\sigma})\mathbf{n}$, where
$\mathbf{n}$ is a unit direction. In fact, Hamiltonian (11) is a
generalization of Rashba model in which
$\boldsymbol{\theta}(\boldsymbol{\pi},\boldsymbol{\sigma})=
\boldsymbol{\sigma}$ [21].

In the usual spin-orbit coupling system, the types of spin-orbit coupling
are $\boldsymbol{\pi}\times \boldsymbol{\sigma}$ and $\boldsymbol{\pi}\cdot %
\boldsymbol{\sigma}$. They will exist as long as the spin-orbit coupling
potential is not weak. However, these two types only present the first level
of spin-orbit coupling, the types which we consider presenting the
complicated level of spin-orbit coupling, such as $\boldsymbol{\pi} \times %
\boldsymbol{\theta}$, can also exist in a spin-orbit coupling
system. Comparing with the first level, the complicated level of
spin-orbit coupling is the coupling of the orbit and the first
level. The first level is proportional to the orbit parameter
$\boldsymbol{\pi}$, and the complicated level is proportional to
the square of the orbit parameter, so one can find that the
complicated level reflects the non-linear character of spin-orbit
coupling system. In fact, there are not only the linear spin-orbit
coupling, but also the non-linear spin-orbit coupling in a
spin-orbit coupling system. The first level of spin-orbit coupling
such as $\boldsymbol{\pi} \times \boldsymbol{\sigma}$ or
$\boldsymbol{\pi} \cdot \boldsymbol{\sigma}$ is the linear
spin-orbit coupling, the complicated level such as
$\boldsymbol{\pi} \times \boldsymbol{\theta}$ reflects the
non-linear spin-orbit coupling. From this point, we say that the
complicated level presents the non-linear character of a
spin-orbit coupling system.

The significance of the topological force and torque lies in the
fact that they provide a complete picture of describing the motion
of a non-relativistic electron in a spin-orbit coupling system. It
is found that the topological force appears as a pure quantum effect
when both electrical and magnetic fields exist in the system, while
the topological torque originates from the topological force and a
gauge potential. Through these force and torque, the
$Zitterbewegung$ effect can be understood deeply, the more
complicated phenomena such as the nonlinear spin-orbit coupling also
can be revealed. This implies that the spin-orbit coupling can
produce more novel effects.

We are grateful to S. Q. Shen, Y. G. Yao and J. R. Shi for helpful
discussions. This work was supported by NSF of China under grant
10347001, 90403034, 90406017, 60525417, 10665003, 10674163, by the
National Key Basic Research Special Foundation of China under
2005CB724508 and 2006CB921400.

\end{document}